\documentclass[%
reprint,
superscriptaddress,
 amsmath,amssymb,
 aps,
]{revtex4-1}

\usepackage{graphicx}
\usepackage{dcolumn}
\usepackage{bm}
\usepackage{xcolor}

\usepackage{mathtools}
\usepackage{ amssymb }

\begin{document}


\title{Collective olfactory search in a turbulent environment}
\author{Mihir Durve$^*$}
 \affiliation{Department of Physics, Universit\`a degli studi di 
Trieste, Trieste, 34127 Italy }
\affiliation {Quantitative Life Sciences, The Abdus Salam International 
Centre for Theoretical Physics - ICTP,  34151,Trieste, Italy}
\author{Lorenzo Piro$^*$}
\affiliation{Department of Physics and INFN, University of Rome Tor Vergata, Via della Ricerca Scientifica 1, 00133, Rome, Italy}
\affiliation{Max Planck Institute for Dynamics and Self-Organization,
Am Fassberg 17,
37077 Goettingen,
Germany}
\author{Massimo Cencini}
\affiliation{Istituto dei Sistemi Complessi, CNR, via dei Taurini 19, 00185 Rome, Italy and  INFN, sez. Roma Tor  Vergata} 
\author{Luca Biferale}
\affiliation{Department of Physics and INFN, University of Rome Tor Vergata, Via della Ricerca Scientifica 1, 00133, Rome, Italy}
\author{Antonio Celani}
\affiliation{Quantitative Life Sciences, The Abdus Salam International 
Centre for Theoretical Physics - ICTP, Trieste, 34151, Italy}

\date{\today}

\begin{abstract}
Finding the distant source of an odor dispersed by a turbulent flow is a vital task for many organisms, either for foraging or for mating purposes. At the level of individual search, animals like moths have developed effective strategies to solve this very difficult navigation problem based on the noisy detection of odor concentration and wind velocity alone. 
 When many individuals concurrently perform the same olfactory search task, without any centralized control, sharing information about the decisions made by the members of the group can potentially increase the performance. But how much of this information is actually valuable and exploitable for the collective task~? Here we show that, in a model of a swarm of agents inspired by moth behavior, there is an optimal way to blend the private information about odor and wind detections with the publicly available information about other agents' heading direction. At optimality, the time required for the first agent to reach the source is essentially the shortest flight time from the departure point to the target. Conversely, agents who discard public information are several fold slower and groups that do not put enough weight on private information perform even worse. Our results then suggest an efficient multi-agent olfactory search algorithm that could prove useful in robotics, for instance in the identification of sources of harmful volatile compounds.

\end{abstract}

\pacs{Valid PACS appear here}
\maketitle

Animals are often on the move to search for something: a food source, a potential mate or a desirable site for laying their eggs. 
In many instances their navigation is informed by airborne chemical cues. One of the best known, and most impressive, olfactory search behavior is displayed by male moths~\cite{david1983,kennedy1983,elkinton1987,dekker2005}. 
Males are attracted by the scent of pheromones emitted in minute amounts by calling females that might be at hundreds of meters away. The difficulty of olfactory search can be appreciated by realizing that, due to air turbulence, the odor plume downwind of the source breaks down into small, sparse patches interspersed by clean air or other extraneous environmental odors ~\cite{murlis1981,celani2014}. The absence of a well-defined gradient in odor concentration at any given location and time greatly limits the efficiency of conventional search strategies like gradient climbing. Experimental studies have in fact shown that moths display a different search strategy composed of two phases: surging, i.e. sustained upwind flight, and casting, i.e. extended alternating crosswind motion. These phases occur depending on whether the pheromone signal is detected or not. This strategy and others have inspired the design of robotic systems for the identification of sources of gas leaks or other harmful volatile compounds~\cite{lilienthal,ferri,ishida,grasso,lochmatter}.  Albeit the effectiveness of individual search is already remarkable in itself, the performance can be further boosted by cooperation among individuals, even in absence of a centralized control~\cite{bonabeau,pitcher,torney2009,ioannou2015,berdahl2013,miller2013}.

In this Letter, we tackle the problem of collective olfactory search in a turbulent environment. When the search takes place in a group, there are two classes of informative cues available to the agents. First, there is private information, such as the detection of external signals -- odor, wind velocity, etc -- by an individual. This perception takes place at short distances and is not shared with other members of the group. Second, there is public information, in the form of the decisions made by other individuals. These are accessible to (a subset of) the other peers, usually relayed by visual cues, and therefore with a longer transmission range. Since the action taken by another individual may be also informed by its own private perception of external inputs, public cues indirectly convey information about the odor distribution and the wind direction at a distance. However, the spatial and temporal filtering that is induced by the sharing of public cues may in principle destroy the relevant, hidden information about the external guiding signals. 

These considerations naturally lead to the question if the public information is exploitable at all for the collective search process. And if it is, how should the agents combine the information from private and public cues to improve the search performances~? Below, we will address these questions by making use of a combination of models for individual olfactory search and for flocking behavior, in a turbulent flow.

\begin{figure} [!ht]
\includegraphics[width=\columnwidth]{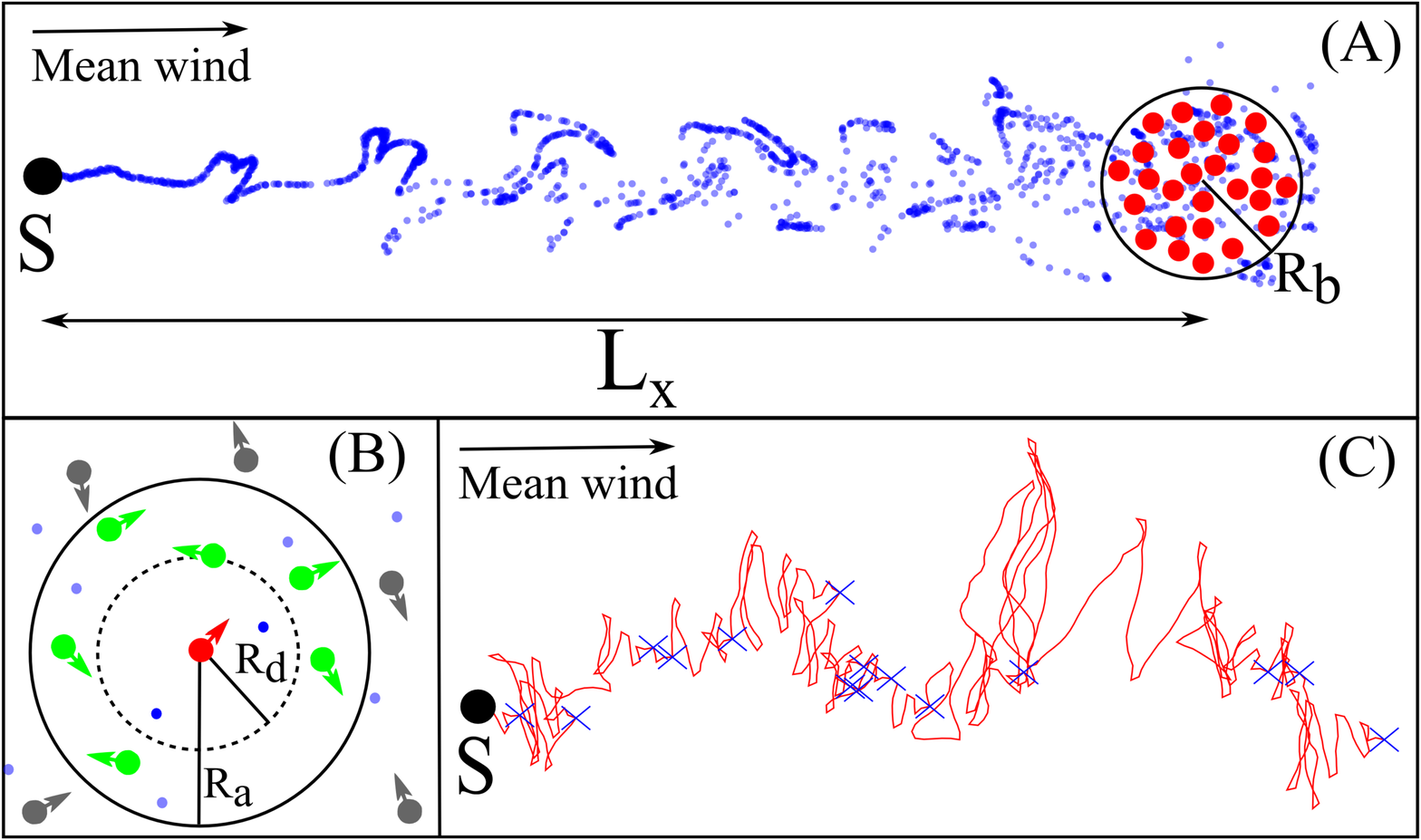}
\caption{\label{fig1}
Collective olfactory search. (A) Odor particles dispersed by the 
turbulent environment are shown by semi-transparent blue dots emitted by the 
source S. Agents (red) are initially  placed far from the source in a 
packed configuration. (B) 
Perception of an agent (red). Detected odor particles 
by the agent are shown as darker blue dots and neighbors of the agent are 
shown in green. Arrows indicate the instantaneous moving direction of agents. We set 
$L_x=250R_d$, $R_b=25R_d$, $R_a=5R_d$, $R_d=0.2$ $b=2.5$ (C) 
Trajectory of an isolated agent performing the cast-and-surge program (see text). The locations where the agent detects the presence of odor particles are shown as blue crosses. }
\end{figure}

\paragraph*{A model for collective olfactory search.}
The setup for our model is illustrated in Fig~\ref{fig1}A. 
Initially, $N$ agents are randomly and uniformly placed within a circle of 
radius $R_b$ at a distance $L_x$ from the source $S$. The odor source $S$ emits 
odor particles at a fixed rate of $J$ particles per unit time. The odor particles are transported in the surrounding environment by a turbulent flow ${\bm u}$ with mean wind ${\bm U}$ (details are given below). 
Notice that the odor particles are not to be understood as actual molecules, but rather represent patches of odor with a concentration above the detection threshold of the agents.  The entire system 
is placed inside a larger square box of size $b L_x$ with reflecting boundary 
conditions for the agents. A complete list of parameters with their numerical values is given in the Supplementary Material.

\paragraph*{Response to private cues.} The behavior elicited by private cues such as odor and wind speed is inspired by the cast-and-surge strategy observed in moths. We adopted a modified version of the 
``active search model"~\cite{balkovsky} that works as follows.

We assume that the agents have access to an estimate of the mean velocity of the wind as moths actually do via a mechanism named optomotor anemotaxis~\cite{baker}. In the model this estimate $\hat{\bm u}(t)$ is an exponentially discounted running average of the flow velocity ${\bm u}$ perceived by the agent along its trajectory: $\hat{\bm u}(t) =\lambda \int_{0}^{t}  \bm{u} (s) \exp[-\lambda(t-s)] ds$. The parameter $\lambda$ is the inverse of the memory time: for $\lambda \to 0$ the estimate converges to the mean wind, while for $\lambda \to \infty$ it reduces to the instantaneous wind velocity at the current location of the agent. In the following we have taken $\lambda=1$ which is of the same order of magnitude of the inverse correlation time of the flow. It is worth pointing out that the only effect of the wind is to provide contextual information about the location of the source. Indeed, in our model the agents are not carried away by the flow, an assumption that is compatible with the fact that the typical airspeed of moths and birds largely exceeds the wind velocity.

At each time interval $\Delta t$, the agent checks if there are odor particles within its olfactory range $R_d$ (see Fig.~\ref{fig1}B). If this is the case, then it moves against the direction of the current estimated mean wind at a prescribed speed $v_0$. When the agent loses contact with the odor cue, it starts the ``casting" behavioral program: it proceeds by moving in a zig-zag fashion, always transversally to the current estimated mean wind, with turning times that increase linearly with the time from the last odor detection (a sample trajectory is shown in Fig~\ref{fig1}C, see the Supplementary Material for details about the implementation). We denote by ${\bm v}_i^{priv}(t)$ the instantaneous velocity of agent $i$ prescribed by this cast-and-surge program. This is uniquely based on private cues and would indeed be the actual velocity adopted by the agent if it were acting in isolation.

\paragraph*{Response to public cues.} To describe the interactions among agents we have drawn inspiration from flocking and adopted the Vicsek model to describe the tendency of agents to align with their neighbors (see \cite{vicsek12,ginelli2016} and references therein). We assume that an individual
can perceive the presence of its peers within a visual range $R_a$ (see Fig.~\ref{fig1}B) and actually measure their mean velocity. According to this model, the behavioral response elicited in agent $i$ by its neighbors is
\begin{equation}
\bm{v}^{pub}_i(t) = v_0 \sum \limits_{j\in{ D_i}} 
{\bm{v}_j}(t) \left/ 
 { \left \vert\left\vert  \sum \limits_{j\in{ D_i}} 
{\bm{v}_j}(t)  \right \vert\right\vert } \right. ,
 \label{vhat}
 \end{equation}
 where $D_i$ is the disk of radius $R_a$ centered around the position of the $i-$th individual. In order to account for errors in the sensing of the velocities of the neighbors we have added, as is customarily done, a noise term in the form of a rotation by a random angle $\bm{v}^{pub}_i(t) \leftarrow R(\theta) \bm{v}^{pub}_i(t)$. Here $\theta$ is independently sampled for each agent and at each decision time from a uniform distribution in $[-\eta \pi,\eta \pi]$. The strength of the noise $\eta$ may range from zero (no noise) to unity (only noise): in the following we set $\eta=0.1$.
 
 In the absence of external cues, and for small enough noise, the group of agents described by this dynamics displays collective flocking and moves coherently in a given direction -- totally unrelated with the source location, however. 

\paragraph*{Combining private and public information.} To study collective olfactory search we then merged the two models above as follows. The velocity of the $i-$th agent is a linear combination of the two prescriptions arising from private and public cues, resulting in the update rule

\begin{equation}
\label{eqn_of_motion}
 \begin{split}
  \bm{v}_i(t) = (1-\beta) \bm{v}_i^{priv}(t) + \beta \bm{v}_i^{pub}(t),\\
  \bm{r}_i(t+\Delta t) = \bm{r}_i(t) + v_0 \frac{\bm{v}_i(t)}{\vert\vert 
\bm{v}_i(t) \vert\vert} \Delta t.
 \end{split}
\end{equation}

The parameter $\beta$, that we have dubbed ``trust'', measures the balance between private and public information. For $\beta=0$ the agents have no confidence in their peers, they ignore the suggestion to align and behave independently by acting on the basis of the cast-and-surge program only. Conversely, for $\beta=1$ agents entirely follow the public cues and discard the private information. 

While it is reasonable to expect that for $\beta=1$ the unchecked trust in public cues leads to poor performances in olfactory search, the nontrivial question here is rather if there is any value at all in public information; that is, in other words, if the best results are obtained for a finite $\beta$ strictly larger than zero.

\paragraph*{Modeling the turbulent environment.} To complete the description of our model, we have to specify the underlying flow and the ensuing transport of odor particles. In  our  simulations  the  flow is given by an incompressible, two-dimensional velocity field, $\bm u(\bm x,t)$ with a constant, uniform mean wind  ${\bm U}$ and statistically stationary, homogeneous and isotropic velocity fluctuations.
The odor particles are considered as tracers whose position, $\bm x$, evolves according to $\dot{\bm x}=\bm u(\bm x,t)$.
For the velocity fluctuations we first considered a stochastic flow and then moved to a more realistic dynamics where the flow obeys the Navier-Stokes equations. 

\paragraph*{Results for the stochastic flow} 
This model flow is characterized by a single length and time scale and is obtained by superimposing a few {Fourier modes whose Gaussian amplitudes evolve} according to an Ornstein-Uhlenbeck process with a specified correlation time. The resulting flow is spatially smooth, exponentially correlated in time and approximately isotropic (see Supplementary Material for details). 

We studied the performance of collective search as a function of the trust parameter $\beta$ while keeping the other parameters fixed to the values detailed in the Supplementary Material.
Initially, the agents are waiting in place without any prescribed heading direction until one 
of the agents detects the odor particles carried by the flow. After this event, agents move as per the equations of motion Eq.~(\ref{eqn_of_motion}). Since the search task is a stochastic process, we run 
many episodes for each value of $\beta$ to compute the average values of several
observables of interest. A given episode is terminated when at least one of the agents is 
within a distance $R_a$ from the source. At this stage we say that the search task 
is accomplished and agents have (collectively) found the odor source. 

We focused our attention on four key observables: {\it (i)} the mean time needed to complete the task which measures the effectiveness of the search; {\it (ii)} the average fraction of agents that, at the time of completion, are close  to the source; {\it (iii)} the order parameter which measures the consensus among members of the group about their heading direction; {\it (iv)}  the degree of alignment of the agents against the mean wind.

\begin{figure} [!h]
\includegraphics[width=\columnwidth]{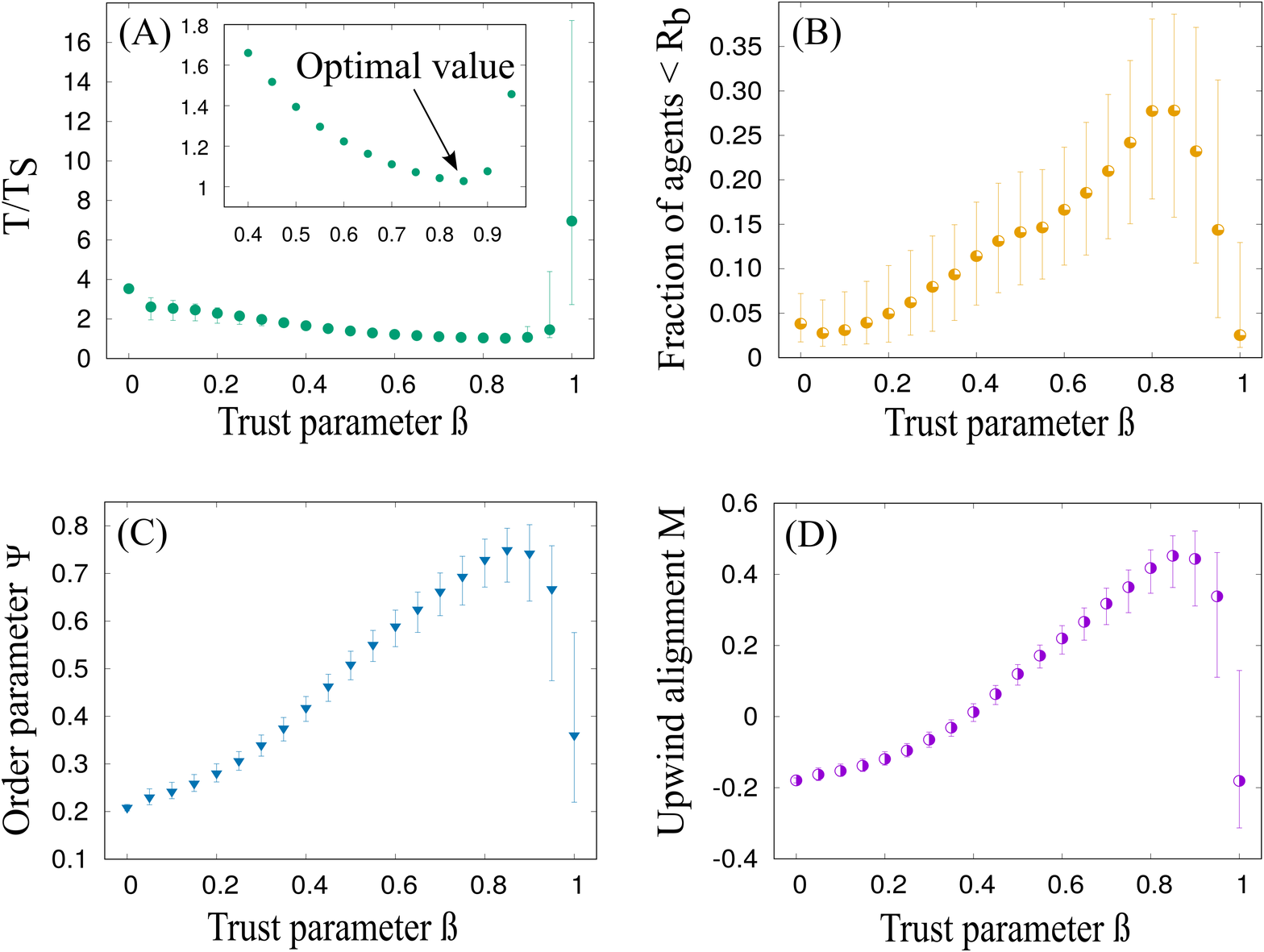}
\caption{\label{result_1}
Collective olfactory search in a stochastic flow. (A) Average search time $T$ for the first agent 
that reaches the target normalized to the straight-path time, $T_s=L_x/v_0$. The 
inset shows a blow-up of region close to the minimum.  (B) Fraction of agents within a 
region of size $R_b$ around the source at the time of arrival of the first agent 
reaching the target. (C) Averaged order parameter $\psi$ (D) Average alignment against the mean wind $M$. For all data, the error bars denote the upper 
and lower standard deviation with respect to the mean value. Statistics is over 
$10^3$ episodes. The parameters were 
set as $\lambda=1$, $N=100$, $J=1$, $\eta=0.1$, $v_0=0.5$, $\Delta t=1$, $L_x=50$.}
\end{figure}

In Fig.~\ref{result_1}A we show the 
average time $T$ for the search completion in units of the shortest path time 
$T_{s}=L_x/v_0$, which corresponds to a straight trajectory joining the target with the center of mass of the flock at the initial time. We observe that there exists an optimal value of the trust parameter 
$\beta \approx 0.85$ for which agents find the odor source in the quickest way. Remarkably, for this value we obtain $ T \simeq 1.03\, T_s$: this means that the agent which arrives first is actually behaving almost as if it had perfect information about the location of the source and were able to move along the shortest path (see movie Beta=0.85.mp4). 

This result has to be contrasted with the singular case of independent agents who act only on the basis of private cues ($\beta=0$) which display a significantly worse performance (the time to complete the task is more than threefold longer)
and move in a zig-zagging fashion (see movie Beta=0.00.mp4). It is also important to remark that the average time grows very rapidly as $\beta$ increases above the optimum. As $\beta$ approaches unity, agents are dominated by the interactions with their neighbors and pay little attention to odor and wind cues. As a result, they form a flock which moves coherently in a direction that is essentially taken at random. If by chance this direction is aligned against the wind, the task will be completed in a short time. However, in most instances the flock will miss the target and either turn because of the noise $\eta$ or bounce on the boundaries until, again by sheer chance, some agent will hit the target (see movie Beta=0.95.mp4). This behavior results in a very long average time accompanied by very large fluctuations. As the outer reflecting boundaries are moved away by increasing $b$, this effect becomes more and more prominent.

\begin{figure*}[!t]
	\centering
	\includegraphics[width=\textwidth]{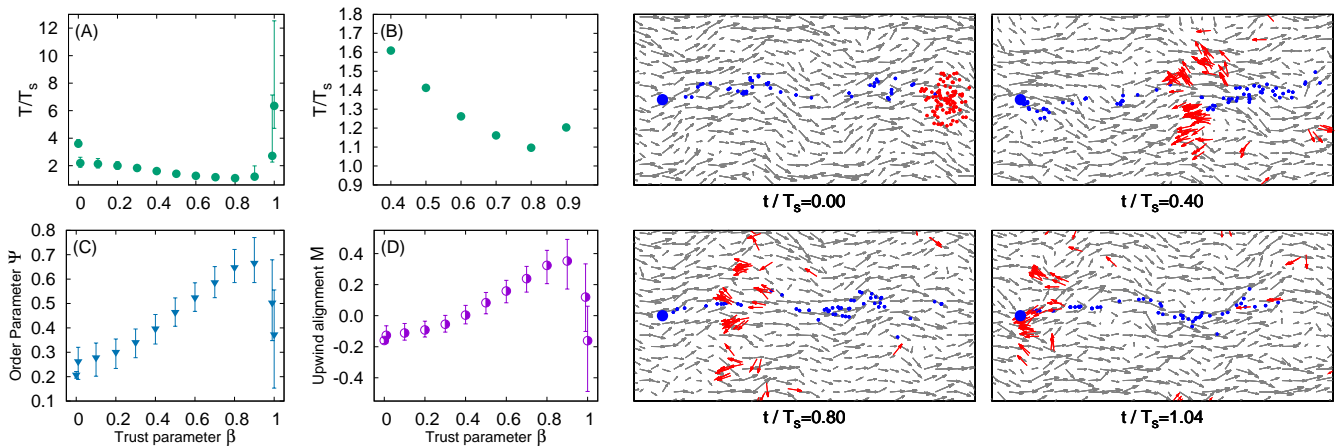}
	\vspace{-0.5truecm}
	\caption{Collective olfactory search in a turbulent flow.
	A: Search time $T$ for the first agent reaching the target normalized by the shortest-path time $T_s$. B: An enlargement of A that highlights the region close to the minimum.  C: Mutual alignment order parameter $\psi$ averaged over time and episodes. D: Average wind alignment $M$.
	Right panels: snapshots of the velocity field (grey arrows) at four different times $t$. The agents (red arrows) navigate in the turbulent flow with the optimal trust parameter $\beta=0.8$. Blue dots represent odor particles dispersed by the flow, while the large blue circle corresponds to the source.   
		\label{fig3}}
\end{figure*}

Since we focused on the time of arrival for the first agent that reaches the source, it is natural to ask what has happened to the other members of the group that have been trailing behind. In Fig.~\ref{result_1}B we show the average fraction of agents that are within a distance $R_b$ (the initial size of the group) when the first agent reaches the target and the task is completed. This quantity is an indicator of the coherence of the group at the time of arrival. It turns out that this fraction has a maximum value $\approx 0.3$ at about the same value of $\beta \approx 0.85$ that gives the best performance in terms of time. This means that on average about one third of the group has been moving coherently along the straight path that connects the initial center of mass of the flock to the target.

To quantify the consensus among agents about which direction they have to take, it is customary to introduce the order parameter
\begin{equation}
 \psi(t) = \frac{1}{Nv_0} \left \vert\left\vert \sum_{i=1}^{N} 
\bm{v}_{i}(t)\right \vert\right\vert  \label{eq:orpar} \; .
 \end{equation} 
When all the agents move in the same direction, whichever it may be, then $\psi=1$. Conversely, if
the agents are randomly oriented then $\psi \simeq N^{-1/2} \ll 1$.
In Fig.~\ref{result_1}C we show the order 
parameter averaged over all agents and all times along the trajectories. As in the previous case we observe a maximum around the range of values of $\beta$ where performance is optimal.

Another parameter of interest is the upwind alignment of the agents 
\begin{equation}
 M(t) = 1- \frac{1}{N} \sum_{i=1}^N \vert\vert \hat{\bm U}  +  \hat{\bm v}_i(t) \vert\vert \label{eq:orpar_m}  \; .
\end{equation}
When all the agents move upwind one has $M=1$ whereas if they all move downwind $M=-1$.
As shown in Fig.~\ref{result_1}D the upwind alignment, averaged over time, 
has a maximum around $\beta=0.85$ which again confirms that a large fraction of the group is heading against the mean wind even if it has access only to a local running time average (the memory time is $\lambda^{-1}=1$, much shorter than $T_s=100$).

The previous results point to the conclusion that there is a relatively narrow range of the trust parameter $\beta$, around $0.85$, for which the collective olfactory search process is nearly optimal, i.e. the time to reach the target is close to the shortest possible one, and takes place with a remarkable coherence of the group.

\paragraph*{Results for a turbulent flow.}
To test the robustness of our findings in a somewhat more realistic situation we also considered the case where the wind velocity is obtained from
 a direct numerical simulation (DNS) of 2D Navier-Stokes equations 
 \begin{equation}
\partial_t  \omega+\bm u \cdot\bm \nabla  \omega= \nu \Delta \omega-\alpha  \omega + f\, ,
\end{equation}
where $\omega =\bm \nabla \times \bm u$, the forcing $f$ acts at small scales so to generate an inverse kinetic energy cascade, that is stopped at large scales by the Ekman friction term with intensity $\alpha$. In order to attain a statistically steady state, the viscous term with viscosity $\nu$ dissipates enstrophy at small scales. In this way we obtain a multiscale flow which is non-smooth above the forcing scale and smooth below it (see \cite{boffettaPRE,boffettaARFM,AB2018} for  phenomenological and statistical flow properties).  DNS have been carried out using a standard $2/3$ dealiased pseudo-spectral solver over a bi-periodic $2\pi\times 2\pi$ box with $256^2$ collocation points and $2^{nd}$ order Runge-Kutta time stepping, see Supplementary Material for technical details. 
In this flow the large scale of the velocity field is about half the size of the simulation box. The numerically obtained velocity field, for a duration of about $10$ eddy turnover times, was then used to integrate the motion of odor particles in the whole plane exploiting the periodicity of the velocity field. Finally, the mean wind is then superimposed. Further details about the simulations are available in the Supplementary Material.

Fig.~\ref{fig3} summarizes the main results obtained with the turbulent flow. As shown in the left panel,
the average time taken by the first agent to reach the source is very similar to the one obtained for the stochastic flow. It displays a minimum time close to the shortest-path time $T_s=L_x/v_0$ at values of the trust parameter $\beta \approx 0.8$. The other observables display very similar features as the ones observed with the stochastic flow.

In the right panels we show four snapshots of the agents at different times during the search process, for $\beta=0.8$, i.e. close to optimality. The flock appears to be moving coherently in the upwind direction and the task is completed in a time $1.04\, T_s$ just a few percent in excess of the nominal minimal time.

\paragraph*{Conclusions and discussion.} We have shown that there is an optimal way of blending private and public information to obtain nearly perfect performances in the olfactory search task. The first agent that reaches the target completes the task by essentially moving in a straight line to the target. This behavior is striking, since in isolation agents move in a zig-zagging fashion (see Fig.\ref{fig1}C). Interestingly, the information about odor and wind is essential to achieve this behavior, but its weight in the decision making is numerically rather small, about 20\%. Although we do not expect that this number stays exactly the same upon changing the various parameters of the model, we suspect that there is a common trend for having optimal values of the trust parameter $\beta$ at the higher end of its spectrum, that is, closer to unity. This may reflect the existence of a general principle of a ``temperate wisdom of the crowds" by which public information must be exploited -- but only to a point. In the present case, one way of summarizing our findings would be the following rule: follow the advice of your neighbors but once every four or five times ignore them and act based on your own sensations. 

With reference to the remarkable similarity between searching in stochastic and turbulent flows shown by Figs. \ref{result_1} and \ref{fig3}, we stress that this is likely due to the specific sensing mechanisms that we have chosen, which is essentially based on single-point single-time measurements. If private cues included consecutive inputs along the agent's trajectory and/or on spatially coarse-grained signals we expect that the results could have been more sensitive to the structure of small-scale and high-frequency turbulent fluctuations.

Our results suggest how to build efficient algorithms for distributed search in strongly fluctuating environments. It is important to point out, however, that our construction is inherently heuristic. Our model heavily draws inspiration from animal behavior, combining features of individual olfactory search in moths and collective navigation in bird flocks. A more principled way of attacking the collective search problem would be to cast it in the framework of  Multi Agent Reinforcement Learning~\cite{busoniu}  and seek for approximate optimal strategies under the same set of constraints on the accessible set of actions and on the available private and public information. It would then be very interesting to see if the strategy discovered by the learning algorithms actually resembles the one proposed here, or points to other known behavior displayed by animal groups, or perhaps unveil some yet unknown way of optimizing the integration of public and private cues for collective search.

\begin{acknowledgments}
M.~D. and L.~P. contributed equally to this work.
L.~B. acknowledges partial funding from the European Union Programme (FP7/2007-2013) grant No.339032. M.~D. is grateful for the graduate fellowship by the ICTP and University of Trieste. M.~D. acknowledges the kind hospitality and support from University of Rome Tor Vergata. M.~D. acknowledges fruitful discussions with A. Pezzotta, M. Adorisio, A. Roy and A. Mazzolini.

\end{acknowledgments}

\bibliography{main}

\end{document}